\documentclass[]{aastex62}

\usepackage{graphicx}
\usepackage{float}
\usepackage[caption=false]{subfig}
\usepackage{enumitem}

\newcommand\aastex{AAS\TeX}

\newcommand{\sha}{K2-22}
\newcommand{\shab}{K2-22~b}
\newcommand{\shb}{Kepler 1520}
\newcommand{\shbb}{Kepler 1520~b}
\newcommand\logten{\ensuremath{\log_{10}}}

\newcommand{\mearth}{M$_\oplus$}
\newcommand{\rearth}{R$_\oplus$}

\newcommand{\mdot}{3~$\times~10^8$ kg/s}
\newcommand{\mdotMePerGyr}{1.6~M$_\oplus$/Gyr}
\newcommand{\lifetimeE}{250~Myr}
\newcommand{\lifetimeIron}{370~Myr}
\newcommand{\lifetimeModel}{21~Myr}
\newcommand{\minRadius}{13~km}
\newcommand{\minRadiusEarths}{0.002~\rearth}
\newcommand{\maxRadius}{4500 km}
\newcommand{\maxRadiusEarths}{0.71~\rearth}
\newcommand{\maxRadiusImprovementFactor}{3.5}
\newcommand{\maxMassEcomp}{2$\times10^{24}$~kg}
\newcommand{\occurrenceProgenitors}{$\sim$0.3\%}

\shorttitle{\aastex\ sample article}
\shortauthors{Schlawin et al.}

\begin{document}

\title{\deleted{Like a Krakatoa Eruption Every Few Minutes: }LBT Reveals Large Dust Particles and a High Mass Loss Rate for K2-22 b}

\correspondingauthor{Everett Schlawin}
\email{eas342 AT EMAIL Dot Arizona .edu}

\author[0000-0001-8291-6490]{Everett Schlawin}
\affiliation{Steward Observatory \\ The University of Arizona \\ 
933 North Cherry Avenue \\
Tucson, AZ 85721, USA}

\author[0000-0002-3532-5580]{Kate Y. L. Su}
\affiliation{Steward Observatory \\ The University of Arizona \\ 
933 North Cherry Avenue \\
Tucson, AZ 85721, USA}

\author{Terry Herter}
\affiliation{Astronomy Department \\
Cornell University\\
Ithaca NY 14853}

\author[0000-0002-5425-2655]{Andrew Ridden-Harper}
\affiliation{Astronomy Department \\
Cornell University\\
Ithaca NY 14853}
\affiliation{Carl Sagan Institute\\
Cornell University\\
Ithaca NY 14853}

\author{D\'aniel Apai}
\affiliation{Steward Observatory \\ The University of Arizona \\ 
933 North Cherry Avenue \\
Tucson, AZ 85721, USA}
\affiliation{Lunar and Planetary Laboratory \\ The University of Arizona \\
1629 E. University Blvd. \\
Tucson, AZ 85721, USA}




\begin{abstract}

The disintegrating planet candidate K2-22 b shows periodic and stochastic transits best explained by an escaping debris cloud.
However, the mechanism that creates the debris cloud is unknown.
The grain size of the debris as well as its sublimation rate can be helpful in understanding the environment that disintegrates the planet.
Here, we present simultaneous photometry with the $g$ band at 0.48~\micron\ and $K_S$ band at 2.1~\micron\ using the Large Binocular Telescope.
During an event with very low dust activity, we put a new upper limit on the size of the planet of \maxRadiusEarths\ or \maxRadius.
We also detected a medium-depth transit which can be used to constrain the dust particle sizes.
We find that the median particle size must be larger than about 0.5 to 1.0~\micron, depending on the composition of the debris.
This leads to a high mass loss rate of about \mdot\ that is consistent with hydrodynamic escape models.
\added{If they are produced by some alternate mechanism such as explosive volcanism, it would require extraordinary geological activity.}\deleted{to produce by explosive volcanism.}
Combining our upper limits on the planet size with the high mass loss rate, we find a lifetime of the planet of less than \lifetimeIron.
This drops to just \lifetimeModel\ when adopting the 0.02~\mearth\ mass predicted from hydrodynamical models.

\end{abstract}

\keywords{stars: atmospheres --- stars: individual (\objectname{K2-22}) ---
stars: variables: general}



\section{Introduction} \label{sec:intro}

The discovery of \shbb /KIC12557548 b \citep{rappaport} ushered in a new class of extrasolar objects that are actively spewing dust off their surfaces.
These disintegrating rocky objects exhibit a variety of different behaviors from periodic, e.g., Kepler-1520 b \citep{rappaport} and \shab\ \citep{sanchis-ojedak2-22}, to having a range of periods, e.g., WD~1145+017 \citep{vanderburg2015wdDisintegrating,gary2017photometricObsWD1145p017} to aperiodic, e.g. KIC~8462852 \citep{boyajian846}.
The dust can cover a significant fraction of the host star and produce transit depths ranging from 0 to $\sim$60\% for the WD~1145+017 system \citep{gansicke2016wd1145p017phot}.
For the \sha\ system, the transit depth range is $\sim$0\% to $\sim$1.3\% \citep{sanchis-ojedak2-22}.
The large variability and $\lesssim$0.14\% minimum in the transit depth indicates that the transits are dominated by dust with no significant absorption from an underlying rocky body.
The dust evaporation is a stochastic process that can vary significantly from one orbit to the next.
So far, there are only upper limits on the underlying planets or asteroids producing the dusty debris.
For \shab, radial velocity measurements constrain the mass  to $<1.4 M_J$ and shallow transit depths constrain the radius to be $<2.5$~R$_\oplus$ \citep{sanchis-ojedak2-22}.

The disintegration mechanism is not fully known, but for the \deleted{disintegrating }systems orbiting main sequence stars like \sha, it is hypothesized to be the hydrodynamic escape of a metal-rich planetary atmosphere that condenses into dust as it cools \citep{perez-becker}.
Another possible mechanism could be explosive volcanism \citep{rappaport}.
There is an apparent correlation between starspot activity and deep transit events in the \shb\ system, which suggests that high energy radiation or magnetic activity may trigger disintegration events \citep{kawahara2013starspots}.
This is because the transit depths of \shbb\ tend to be 30\% deeper when the stellar flux is below average for \shb\ (when the hemisphere facing Earth is covered by more spots) \citep{kawahara2013starspots}.
However, there is an alternative suggestion that starspot occultations by the dusty debris and random groupings of starspots could create an apparent correlation \citep{croll2015starspots}.
The 82 days of observations of \sha\ do not have the precision necessary to explore this same correlation between stellar flux and transit depths \citep{sanchis-ojedak2-22}.

An explosive volcanic model has not been explored in detail, but \citet{perez-becker} calculate the mass loss history and estimate plausible masses that are compatible with hydrodynamic escape.
The mass loss rate increases until it reaches a catastrophic phase toward the end of the planet's lifetime and in this free-streaming limit the mass loss reaches a plateau.
\citet{perez-becker} predict that there may be a population of progenitor Mercury-sized planets that are below the detection threshold of current planet searches.
\added{The hydrodynamic escape model also explains on-off behavior observed in the \shb\ system where the transit depths sometimes alternate between deep and shallow transit events with each orbital period \citep{rappaport,vanWerkhoven2014}.
This is explained by a limit cycle where the deep transit events cause shadowing and cooling of the surface and can be followed by low evaporation and shallow transit depths.
These shallow transit depths expose the surface to irradiation and thus enhance atmospheric escape to produce deep transit depths later \citep{perez-becker}.
We note that the hydrodynamic escape model is driven by visible-wavelength radiation and not X-rays or ultraviolet, so there is no requirement that stellar activity and deep transit events are correlated in the hydrodynamic escape model.
}

Disintegrating rocky planets provide laboratories to study the solid materials in other planetary systems.
This complements studies of white dwarf atmospheric composition, where the stellar material is polluted with material from surrounding asteroids or planets \citep{jura2003wdPollution}.
So far, white dwarf atmospheric studies have revealed accurate abundances of solid material and that the accreted planetesimals largely match the bulk elemental composition of the Earth and other rocky bodies \citep{jura2014cosmochemistry,zuckerman2018wdPlanetMaterialReview}.
However, disintegrating bodies orbiting main sequence stars provide an avenue to studying planets before the giant stages and death of a star.

Future mid-infrared to far-infrared spectroscopic observations could yield the composition of the debris of the escaping dust.
The dust debris can then be traced to either the core, mantle or crust material of a terrestrial planet.
Quartz could indicate crust material, silicates could indicate mantle material, while iron can indicate core material \citep{bodman2018disintegratingInteriors,okuya2020spicaDisintegratingComposition}.
If the planet is a coreless body \citep{elkins-tanton2008coreless}, it may also produce substantial crystalline fayalite (Fe$_2$SiO$_4$) dust \citep{okuya2020spicaDisintegratingComposition}.
It is likely that the evaporating metal rich atmosphere around a planet will evaporate as much gas as dust in the system \citep{perez-becker}.
So far, sodium gas has been ruled out at $\lesssim$10\% and ionized calcium gas at $\lesssim$2\% at 5$\sigma$ confidence using high resolution spectroscopy \citep{ridden-harper2019searchForGasK2d22}.

It is possible to indirectly constrain the composition of the dust debris by matching sublimation timescales with the observed tail length and particle size distribution \citep{vanlieshout2014kic1255comp,vanlieshout2016kic1255}.
\citet{ridden-harper2018chromaticKIC1255} show that the variable transmission spectrum of \shbb's dust tail can be explained by varying optical depth and vertical extent.
We also note that the debris's vertical extent can also be a function of ejection speed and planet mass, so observations can constrain those related variables.
In contrast to the trailing tail observed with \shbb, \shab\ has a leading tail, likely due to a lower ratio between radiation pressure forces and gravitational forces.

The particle size distribution can be constrained by the transmission spectrum of the debris because of the wavelength-dependence of dust extinction \citep{bochinski2015evolving,schlawin2016kic1255,croll2014,sanchis-ojedak2-22}.
Multi-band transit photometry can also be used to constrain the particle sizes of the dust escaping disintegrating planets because the dust extinction has a broad shape.
\citet{colon2018groundK2-22campaign} observed \sha\ with a large ground-based campaign and find that there is no significant color dependence to the transits across 0.5~\micron\ to 0.9~\micron\ for high precision simultaneous coverage.
\citet{colon2018groundK2-22campaign} find suggestive evidence that the $g$ band transit depths are larger on average than the $i$ band from stacked lightcurves but there were differences in precision and observing windows so it is not possible to make a definitive detection of wavelength dependence other than the Gran Telescopio CANARIAS (GTC) observations of a single deep transit on February 14, 2015 \citep{sanchis-ojedak2-22}.

Here, we obtain multi-wavelength simultaneous lightcurves using the Larger Binocular Telescope to study the debris escaping from \shab.
High precision multi-wavelength lightcurves can provide valuable information about \shab's particle size distribution, disintegration mechanism and indirectly provide information about composition.

LBT imaging with different filteres on the two 8.4~m mirrors is particularly suited to disintegrating planet observations because it is critical that the observations in different filters occur simultaneously.
The amount of escaping debris from a disintegrating planet is highly stochastic and unpredictable from one planet orbit to the next.
If observations are made in one filter on one night and a different wavelength filter a few orbits later, there is no way to distinguish the filter-dependence from temporal variability.
Simultaneous lightcurves in different photometric bands, however, reveal the wavelength dependence alone.

Section \ref{sec:obs} describes our multi-wavelength observations.
We fit the lightcurves in Section \ref{sec:lcFit}.
These lightcurves are used to constrain the particle size distribution in Section \ref{sec:rDistribution}.
We use one of the lightcurves that shows a very shallow transit depth to constrain the size of the underlying disintegration planet and its lifetime in Section \ref{sec:planetSize}.
We conclude in Section \ref{sec:conclusions}.

\section{Observations}\label{sec:obs}

We observed the \sha\ system with the Large Binocular Telescope \citep{rothberg2018lbtInstruments} on three partial nights.
We used the heterogenous binocular mode which allows the observatory to simultaneously point to the same target using the large binocular camera (LBC) \citep{giallongo2008LBC,speziali2008LBC} on the SX (Left) 8.4~m mirror and the LBT Utility Camera in the Infrared 2 (LUCI2) instrument on the DX (Right) 8.4~m mirror.

We chose observing modes with the shortest and longest possible wavelengths where high signal to noise ($>$3000 in 10 minutes) is achievable, providing the largest change in the dust extinction function and thus the best constraints on the particle sizes.
The filters satisfying this criterion are the $g'$ band for the LBC blue instrument with a midpoint of 0.47~\micron\ and the $K_S$ band on the LUCI2 instrument with a midpoint of 2.16~\micron.
We selected the $K_S$ filter over the $K$-band filter to minimize the effects of telluric contamination on the observations.
Imaging and spectroscopy are both possible with LUCI2, but we elected to use an imaging mode to minimize the possible sources of systematic error or slit loss.

For the LBC $g'$ observations, we followed the modes and observing strategy of \citet{nascimbeni2013gj3470bLBT} and \citet{nascimbeni2015gj1214LBT}, which successfully achieved high precision observations of transiting exoplanet systems.
The LUCI2 images require many calibration steps to decrease the noise in the lighcurves.
All detector images must be corrected for non-linearities.
For the detector non-linearity correction we use the simple (1 parameter) quadratic correction from the LBTO \footnote{https://sites.google.com/a/lbto.org/luci/instrument-characteristics/detector}.
The flat field image calibration also required many steps, which are described in Appendix \ref{sec:flatFieldCreation}.

\begin{deluxetable*}{ccCrlc}[b!]
\tablecaption{Summary of observations\label{tab:observations}}
\tablecolumns{6}
\tablewidth{0pt}
\tablehead{
\colhead{UT start date} &
\colhead{K Band Focus} &
\colhead{Seeing} & \colhead{Filters} & \colhead{Transit Description} \\
\colhead{(YYYY-mm-dd)} & \colhead{} &
\colhead{(arcsec)} & \colhead{} & \colhead{}
}
\startdata
2020-01-25 & In-focus & 1.6 - 1.7 & $g,K_S$ & Partial Transit \\
2020-01-28 & De-focus & 1.6 - 3.5 & $g, C\tablenotemark{a}, K_S$ & Medium \\
2020-02-20 & In-Focus & 0.8 - 1.0 & $g,K_S$ & Shallow \\
\enddata
\tablenotetext{a}{Open CCD/broadband.}
\end{deluxetable*}

The Hereford Arizona Observatory (HAO), with its 16-inch Ritchey-Chr\'etien telescope, was also used to monitor lightcurves of \sha\ simultaneously during two of the LBT campaigns.
Given the faintness of the target, HAO photometry was obtained with an unfiltered CCD, thus integrating light from 0.4~$\mu$m to 0.85~$\mu$m, making it similar to $Kepler$/$K2$'s bandpass but with wider wings.
HAO is located at 1,400 m altitude, so telluric absorption is also imprinted on the intrinsic instrument bandpass.
The observing setup and reduction method is the same as \citet{rappaport2016wd1145p017driftingFragments}.
This photometry was used as an intermediate bandpass between the $g$ and $K_S$ band.

\begin{figure*}
\gridline{
	\fig{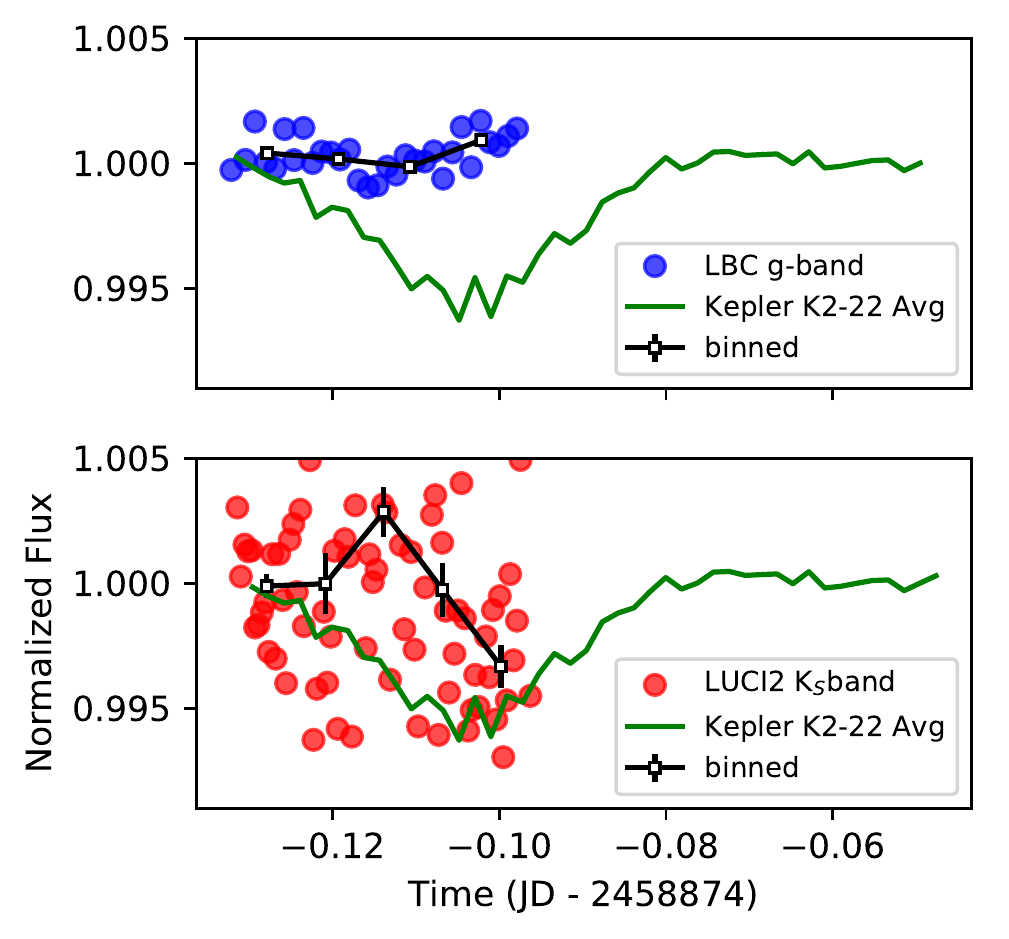}{0.49\textwidth}{UT 2020-01-25}
	\fig{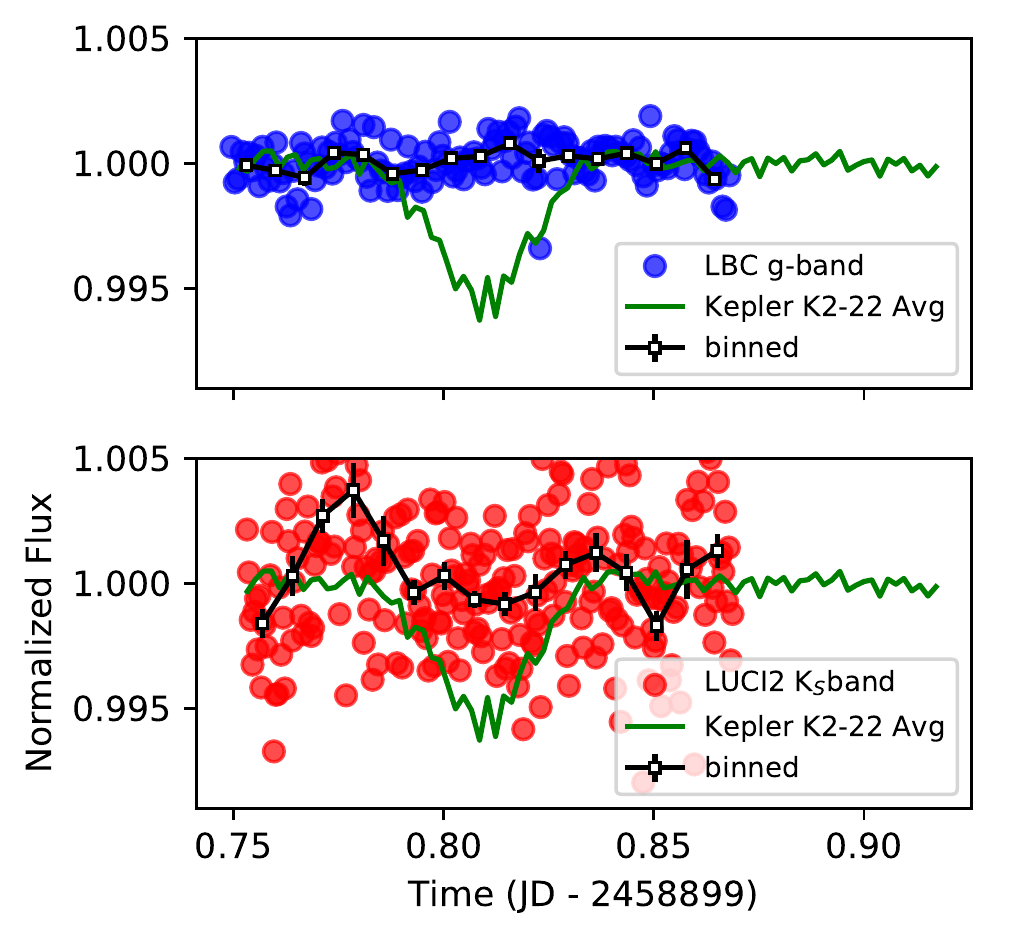}{0.49\textwidth}{UT 2020-02-20}
	          }
\caption{Photometry from UT 2020-01-25 and UT 2020-02-20 where transit depths were shallower than the K2 82 day average.
The raw photometry is shown as blue and red points for the g band and $K_S$ band, respectively, with the binned data as white squares with black error bars.
The average phase-folded K2 lightcurve is shown as a green line.
The photometry from UT 2020-02-20 (right) is used to constrain the size of the planet.
}\label{fig:photJan25Feb20}
\end{figure*}

\begin{figure*}
\gridline{
          \fig{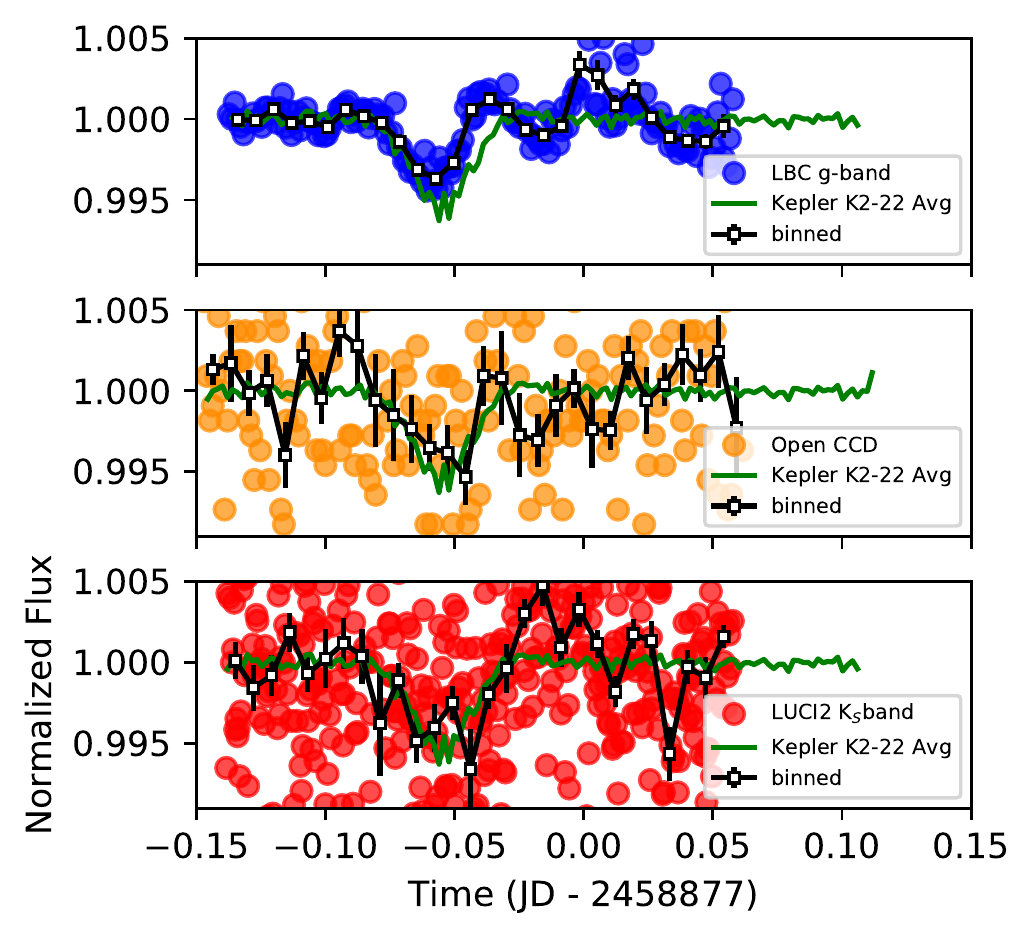}{0.49\textwidth}{UT 2020-01-28}
          \fig{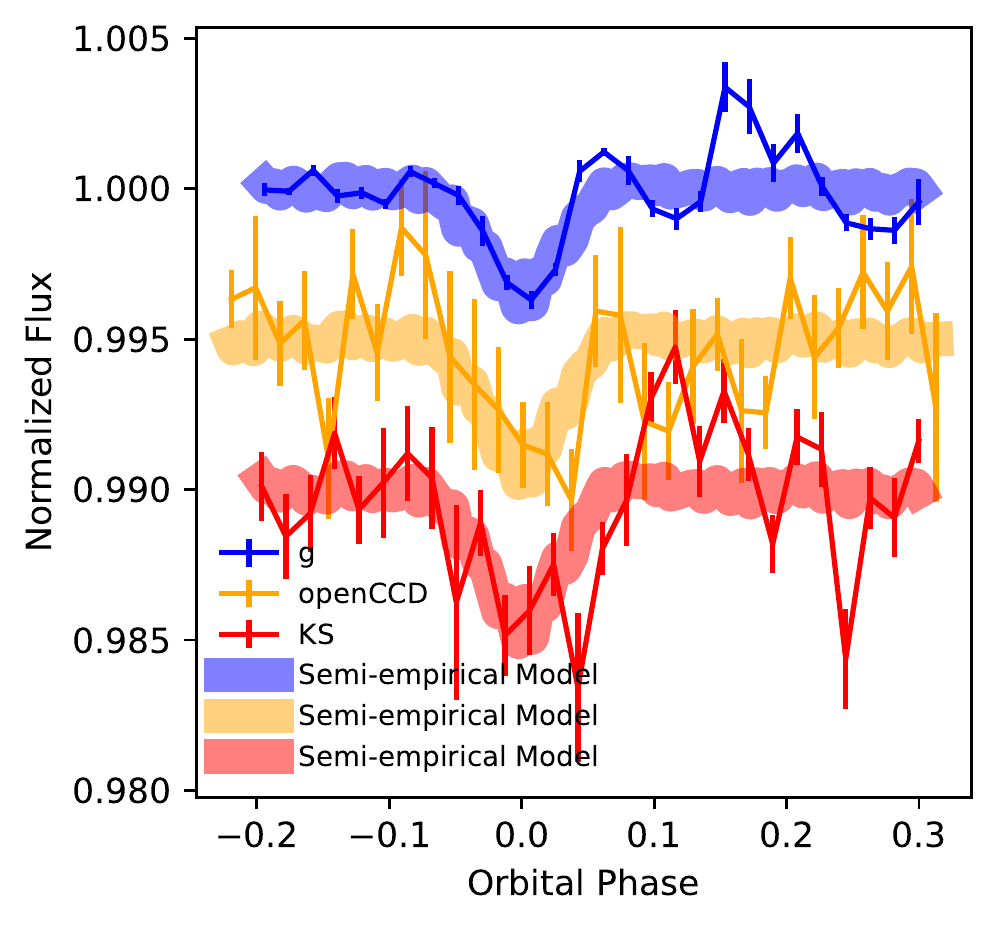}{0.49\textwidth}{UT 2020-01-28, binned}
          }
\caption{Photometry from UT 2020-01-28, where there was a medium-depth transit.
The 82 day average lightcurve from K2 is scaled to best-fit the data, and the comparison of this scaled model with binned data is shown on the right.
}\label{fig:photJan28}
\end{figure*}

A summary of the three nights of observations is listed in Table \ref{tab:observations}.
To plan the observations and calculate baselines before and after transit, we use the ephemeris from \citep{sanchis-ojedak2-22}, which has an uncertainty of 8 minutes on UT2020-01-28.
Figure \ref{fig:photJan25Feb20} shows photometry from UT2020-01-25.
The first observation did not include enough photometry preceding the ingress of disintegrating debris to sufficiently measure the pre-ingress baseline, so this night was used as a test of the instrument mode in preparation for future observations.
On UT2020-01-28, \deleted{the} \shab\ exhibited a moderately deep transit, as shown in Figure \ref{fig:photJan28}.
Finally, on UT2020-02-20, \shab\ exhibited a very shallow transit that was consistent with the out-of-transit baseline (Figure \ref{fig:photJan25Feb20}).
This very shallow transit is used to constrain the size of the planet.
For UT2020-01-28 and UT2020-02-20, we removed a linear baseline trend from the data, due either to stellar variability or a systematic drift error.
The lightcurves show the time series after removing this linear trend.
We bin the lightcurves to better visualize the transit depths in Figure \ref{fig:photJan25Feb20} and Figure \ref{fig:photJan28}.
The time bins are separated by 10 minutes and the errors are estimated from the standard deviation of the mean of the points within a bin.

We compare our photometry to the average level measured by the Kepler observatory, which observed \shab\ for 82.2 days.
As can be seen in Figure \ref{fig:photJan28} for 2020-01-28, the $K_S$ band transit depth was consistent with the average from the K2 mission, whereas the $g$ band was slightly shallower.
On 2020-02-20, both photometric bands exhibit shallow transit depths compared to the K2 average (see Section \ref{sec:planetSize} for more details on this lightcurve).

The previous ephemeris has an uncertainty of 8 minutes at the time of observations.
We do find that the g' and K' lightcurves for 2020-01-28 are best-fit with our model with a 7 minute shift to the data (after performing barycentric timing corrections).
We do note that the lightcurve shape and timing can be variable by up to 5-10 minutes as measured by the GTC \citep{colon2018groundK2-22campaign}.
Thus, we cannot robustly update the ephemeris with a single transit event.
We report a tentative new ephemeris with $T_0$ = 2458876.940 $\pm$ 0.001 BJD and P = 0.3810771 $\pm 2 \times 10^{-7}$ d, assuming that the transit shape for UT 2020-01-28 is close to the average from the $K2$ mission.
We adopt this ephemeris for the rest of the paper for lightcurve analysis but stress that this ephemeris should be revised by an average of many transit profiles in future observational campaigns like \citet{colon2018groundK2-22campaign}.

\section{Constraints from Lightcurves}\label{sec:lcConstratins}
The transit depths allow us to derive constraints on the particle sizes and the planet radius from the lightcurves.
We first use the lightcurves where the transit depth was similar to the average from the K2 measurements to study the spectrum of the dusty debris (Section \ref{sec:lcFit}).
We then model this spectrum with a dust distribution (Section \ref{sec:rDistribution}).
Next, we use the lightcurves where the transit depth was very shallow to constrain the planet radius (Section \ref{sec:planetSize}).

\subsection{Lightcurve fitting}\label{sec:lcFit}

We used the average K2 measurements as a model to fit our measured lightcurves because the K2 lightcurves generally follow the same ``V'' shape.
With a few higher precision ground-based GTC lightcurves \citep{colon2018groundK2-22campaign}, there are some shape variations, but we assume that the shape follows the average from the Kepler K2 measurement to reduce the number of free parameters in the tail geometry and because there can be time-correlated errors in ground-based lightcurves that can mimic shape variations.
The model assumes that the dust tail leading \shab\ varies only in optical depth (while staying optically thin with $\tau << 1$.).
We note that kinematic models of the dust ejected at $\sim$1 km/s predict that the dust particles spread out to an optically thin regime when they reach a distance more than a few $R_\oplus$ from the planet \citep{vanlieshout2016kic1255}. 
Thus, we model the lightcurve $f(\phi)$ as a function of orbital phase $\phi$ as 
\begin{equation}\label{eq:phaseModel}
f(\phi) = (f_{K2}(\phi) - 1.0) a + 1.0,
\end{equation}
where $f_{K2}(\phi)$ is the average K2 lighturve and $a$ is a unitless scale factor describing the transit depth.
The resulting best fit lightcurves are shown in Figure \ref{fig:photJan28} (right).

We use the \texttt{python} function \texttt{scipy.optimize.curve\_fit} to fit the lightcurves.
The resulting scale factors in the $g$ band, $C$ band and $K_S$ bands are $a_g$ = 0.63$\pm$0.09, $a_C=0.76 \pm0.21$  and $a_{K_S}$ = 0.81 $\pm$ 0.25, respectively.
Thus, the transit depth in the $K_S$ band is greater or equal to the $g$ band, which indicates that the dust particles are of order-of-magnitude as large as the longer of the wavelengths (2.16~\micron).
In the next section, we will use a quantitative model to explore the grain size distributions in further detail.

\subsection{Inferred Particle Size Distribution}\label{sec:rDistribution}

We next use the transit depths that were fit in Section \ref{sec:lcFit} to put constraints on the particle size distribution of the dusty debris escaping \shab.
We assume that the dust cloud is optically thin so that the transit depth scale factor ($a$) is directly proportional to the average dust extinction function $\bar{Q}_{ext}$.
Our model is then
\begin{equation}
a(\lambda) = A \bar{Q}_{ext}(\lambda),
\end{equation}
where $a$ is the transit depth scale factor from equation \ref{eq:phaseModel}, $A$ is a scale factor that combines the geometric coverage, particle cross section as well as the column density and $\bar{Q}_{ext}(\lambda)$ is the extinction function of the material at a given wavelength, averaged over all particles.
\begin{equation}\label{eq:avgQ}
\bar{Q}_{ext}(\lambda) = \frac{\int_{0}^{\infty} n(r) Q_{ext}(\lambda, r) dr }{\int_{0}^{\infty} n(r) dr},
\end{equation}
where $r$ is the particle radius and $n(r)$ is the number of particles with that radius.
We start by assuming that the composition of the debris is crystalline forsterite (Mg$_2$SiO$_4$) and that the dust is composed of spherically homogeneous grains.
We use the complex indices of refraction for forsterite \citep{suto2006crystallineSilicates} and calculate $Q_{ext}$ using \texttt{miepython}\footnote{https://miepython.readthedocs.io/en/latest/}.
Later, we repeat the analysis for a suite of possible compositions.

We next use \texttt{emcee} \citep{foreman-mackey2013emcee} to derive constraints on the particle size distribution of the dusty debris escaping \shab.
We assume a log-normal particle size distribution \replaced{or}{so that the} number of grains $n(r)$ as a function of radius $r$ follows
\begin{equation}
n(r) = \frac{1}{r \sigma \sqrt{2\pi}}\exp\left( - \frac{(\ln(r) - \ln(r_{med}))^2}{2\sigma^2} \right),
\end{equation}
where $r_{med}$ is the median size of the grains and
$\sigma$ is the logarithmic width of the distribution.
We fix $\sigma$ at 0.5 to create a spread of particles that span several factors of $r_{med}$ below and above the median radius.
Thus, the only free parameters are $r_{med}$ (measured in microns) and the geometric scale factor $A$, so the goal is to find a posterior distribution constrained by the data.
A log-normal distribution is commonly used to describe atmospheric aerosols \citep{heintzenberg1994lognormalParticleSizes} and Brownian coagulation produces aerosols with an approximately log-normal distribution \citep{friedlander1977smokeDustHaze}.
A log-normal particle size distribution is found for amorphous pyroxene grains after vaporizing natural crystalline enstatite in a laboratory and measuring the small grain sizes \citep{brucato1999annealedPyroxene}.
We use 128 points to evaluate $n(r)$ that are spaced equally in logarithmic space and above a threshold of $10^{-3}$ times the total number of particles to approximate the integral in equation \ref{eq:avgQ}.

\begin{figure*}
\gridline{\fig{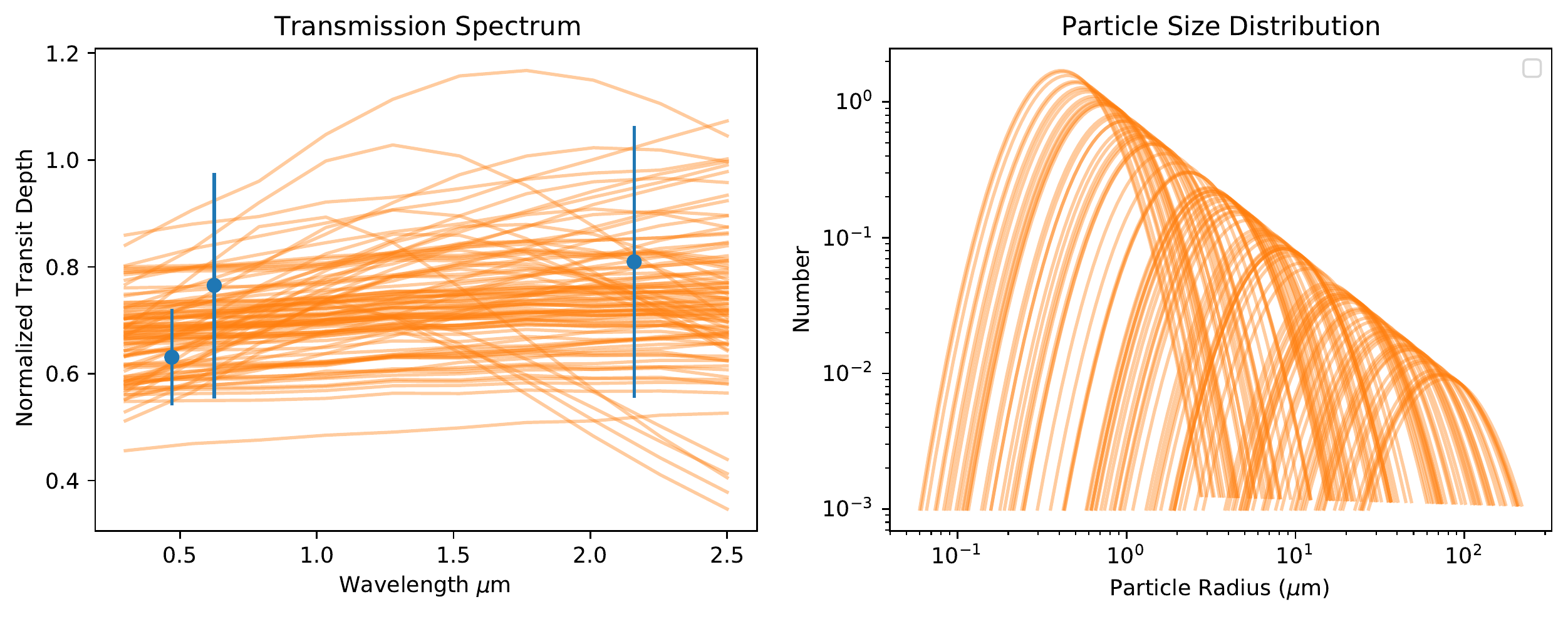}{0.98\textwidth}{}}
\caption{
{\it Left:} Transit depths from UT 2020-01-28 (blue points with error bars) as compared to Mie extinction models for forsterite grains.
Model spectra for 100 random MCMC samples are shown in orange.
{\it Right:} Particle Size distribution for the same set of MCMC samples.
}\label{fig:sedFitsJan18}
\end{figure*}

We assume a flat prior for the logarithm of the median particle size between $\logten(r_{med}) = -4$ to $\logten(r_{med})=2$.
We use 5 \texttt{emcee} walkers and run the chains for 2,000 to 3,000 steps to ensure that the chains have reached at least 50 times the autocorrelation time.
The number of steps needed was different for different dust compositions.
We estimate the maximum autocorrelation time for all variables with \texttt{emcee} and discard 5 times this time to remove the burn-in phase of the MCMC.
Figure \ref{fig:sedFitsJan18} (left) shows \deleted{the maximum likelihood spectrum as well as} the spectra for 100 random samples of the posterior for $r_{med}$ and $A$ for forsterite composition.
Figure \ref{fig:sedFitsJan18} (right) shows the particle size distribution for the \deleted{maximum likelihood and} 100 random samples.
It is clear that large particle sizes $r_{med} \gtrsim 0.5$\micron\ are favored by the measured photometric depths.

\begin{figure*}
\gridline{\fig{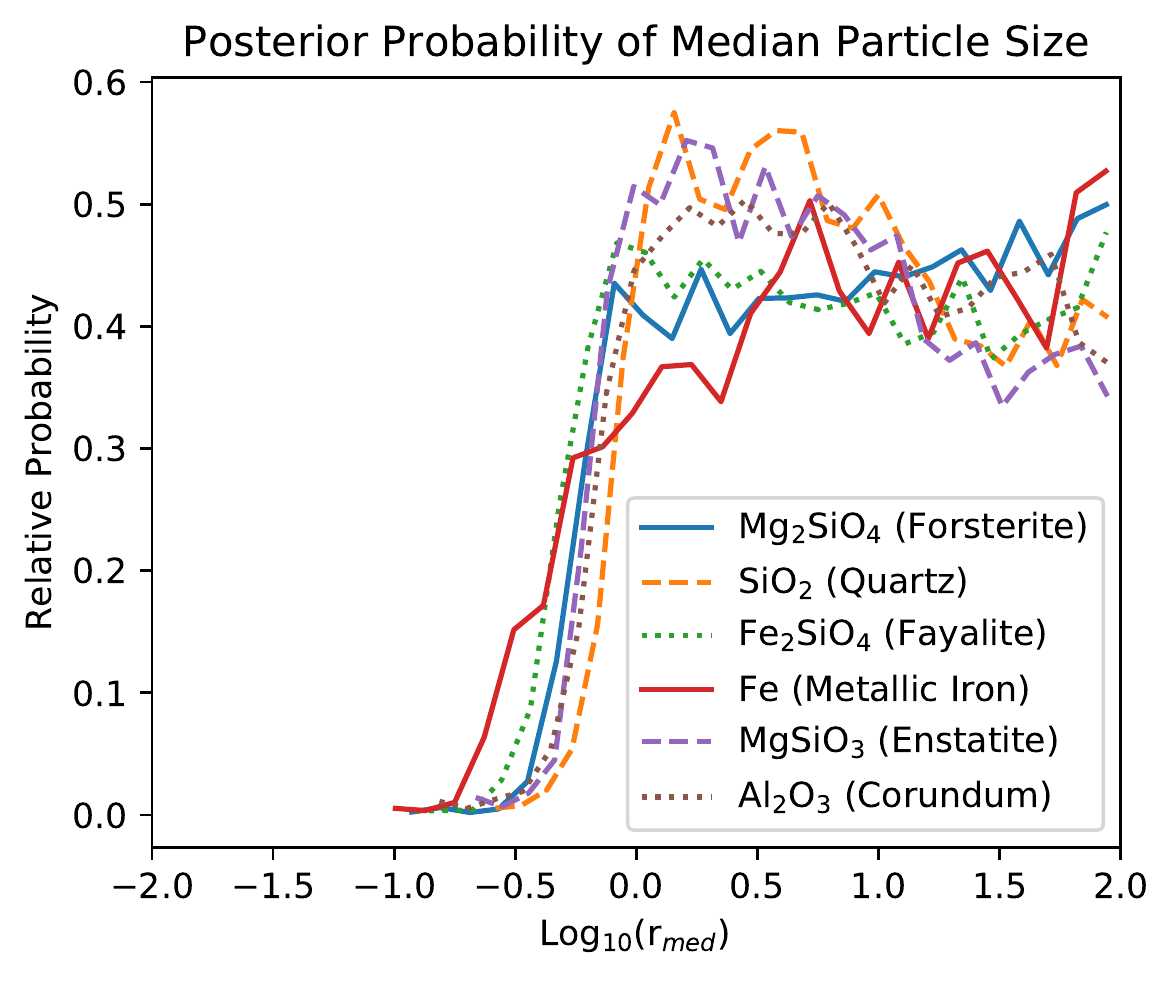}{0.48\textwidth}{}}
\caption{Posterior distribution of the median radius of the particle size distribution for different compositions.
The 5\% posterior lower limit of the median particle radius is 0.5~\micron\ for iron to 0.9~\micron\ for quartz.}\label{fig:posteriorRadiusJan18}
\end{figure*}

In Figure \ref{fig:posteriorRadiusJan18}, we show the posterior distribution for the median particle size of the distribution.
95\% of the highest posterior density is above 0.76~\micron\ for forsterite.
We repeated this posterior distribution fit for SiO$_2$ (quartz), Fe (metallic iron), Fe$_2$SiO$_4$ (fayalite), MgSiO$_3$ (enstatite) and Al$_2$O$_3$ (corundum).
Our code to evaluate the extinction coefficients is available as a Python package\footnote{https://dust-mie.readthedocs.io/en/latest/}.
The 5\% lower limit for the median dust particle size of the distribution ranges from 0.5~\micron\ for iron to 1.0~\micron\ for quartz.
We therefore need relatively large particle sizes for \shab\ for all compositions.

The particle sizes allow estimates of the minimum mass of the material escaping \shab.
We estimate the transit mass as
\begin{equation}
m_d \gtrsim f_{depth} R_*^2 \frac{4}{3} \pi r_{med} \rho_{dust},
\end{equation}
where m$_d$ is the total mass of the dust, $f_{depth}$ is the average transit depth, $R_*$ is the stellar radius, $r_{med}$ is the median particle size lower limit and $\rho_{dust}$ is the density of dust particles \citep{perez-becker}.
Here, we adopt the stellar radius from \citet{sanchis-ojedak2-22} of 0.57 $\pm$ 0.06 R$_\odot$ and an average transit depth of $f_{depth} = 0.55\%$.
If we assume a 0.5 to 0.9~\micron\ particle radius for iron and quartz respectively, a density of 7.9 and 2.6 g/cm$^3$ \citep{vanlieshout2016kic1255}, then the total mass of the dust is $m_d \approx$ 1.3~$\times10^{13}$ kg and 9.3~$\times10^{12}$ kg for iron and quartz respectively.
This assumes that the dust is optically thin with a cross section of $\pi r_{med}^2$.
The material can appear and disappear on orbital timescales so we estimate the mass loss rate as
\begin{equation}
\dot{M} = \frac{m_d}{P_p},
\end{equation}
where $P_p$ is the period of the planet \citep[0.381078 days,][]{sanchis-ojedak2-22}.
Thus we find $\dot{M} = 4\times10^{8}$~kg/s and \mdot\ for iron and quartz respectively.
This is \deleted{a} 2 to 2.7 \added{times} larger than the $1.5 \times 10^8$~kg/s estimate in \citet{sanchis-ojedak2-22} likely because of the larger dust size reported in this work.
We note that the compositional dependence is small compared to other factors such as the particle size distribution function, optical depth and clearing timescales.
We therefore adopt \mdot\ as the lower limit for the characteristic $\dot{M}$.
The true mass loss rate can be significantly higher than this value if it is optically thick or composed of particles that are much larger than 0.5~$\micron$, which is the lower limit on the median particle size estimated above.
These mass loss rates only include the solid particles and the gas loss rate may be a factor of several larger considering that the gas is responsible for the hydrodynamic escape and may entrain dust particles.

\added{
We also compute $\dot{M}$ for the \shb\ system, which has similar properties.
The transmission spectrum is also flat from the optical to 2.2~\micron\ \citep{croll2014}, giving a median particle radius $r_{med} \gtrsim 0.5$~\micron\ \citep{schlawin2016kic1255}.
The average transit depth is nearly the same as \shab\ at 0.6\% \citep{rappaport} and the radius of the star is larger at 0.69~$R_\odot$ \citep{schlawin2018kic1255Normal}.
The period is longer at 0.653 days \citep{rappaport}.
\shab\ has a slightly larger estimate of $\dot{M}$ of \mdot\ versus $1.3 \times 10^8$ kg/s for \shbb\ due to its shorter orbital period and larger median particle radius inferred from this work for a quartz composition.
Altogether, the minimum mass loss rate for \shbb\ and \shab\ are consistent within the range of assumptions like spherical homogeneous grains, averages over time and optical thinness.
}

\deleted{For comparison with the most volcanically active body in our solar system, Io expels matter at a rate of $\sim$1000 kg/s \citep{geissler2003volanicActivityIo}.
For comparison with terrestrial volcanoes, the mass of aerosols emitted by the Krakatoa 1883 volcanic eruption is estimated to be 3$\times 10^{10}$~kg \citep{rampino1982historicVolcanicEruptions}.
Therefore, the $\dot{M}$ we estimate for \shab\ is similar to a Krakatoa 1883 volcanic eruption (counting the aerosols only) about every 100 seconds or 1.7 minutes.
The very large $\dot{M}$ for \shab\ therefore requires a hydrodynamic escape mechanism \citep{perez-becker} or orders of magnitude higher volcanic activity than has been seen in the Solar System.
Finally, another terrestrial comparison is Niagara Falls, which has a modern flow rate of 1770 m$^3$/s \citep{hayakawa2009niagaraFallsRecession} so $\dot{M}$ is 170 times this rate.}
\added{The mass loss rate of \mdot\ for \shab\ is consistent with a hydrodynamic escape mechanism, which produces mass loss rates $\gtrsim 10^8$ kg/s \citep{perez-becker}.
Another possible escape mechanism that has been proposed is explosive volcanism \citep{rappaport}.
However, \mdot\ is orders of magnitude higher than the most volcanically active body in our solar System, Io, which expels matter at a rate of $\sim$ 1000 kg/s \citep{geissler2003volanicActivityIo}.}
We return to the mass loss and estimates of the planet lifetime after finding constraints on the planet size.

\subsection{Planet Size Limits}\label{sec:planetSize}
\begin{figure*}
\gridline{
          \fig{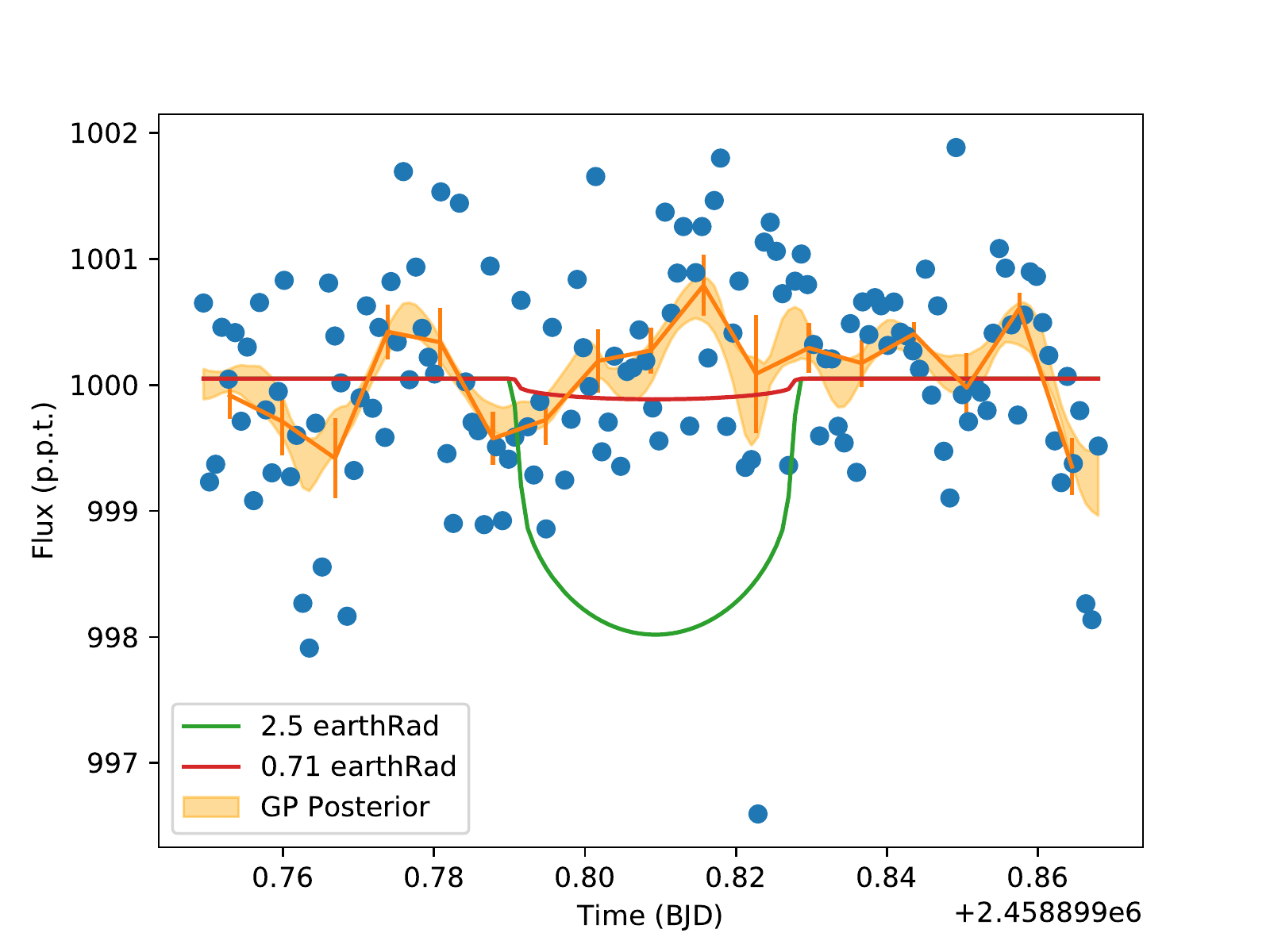}{0.49\textwidth}{Lightcurve from UT 2020-02-20}
          \fig{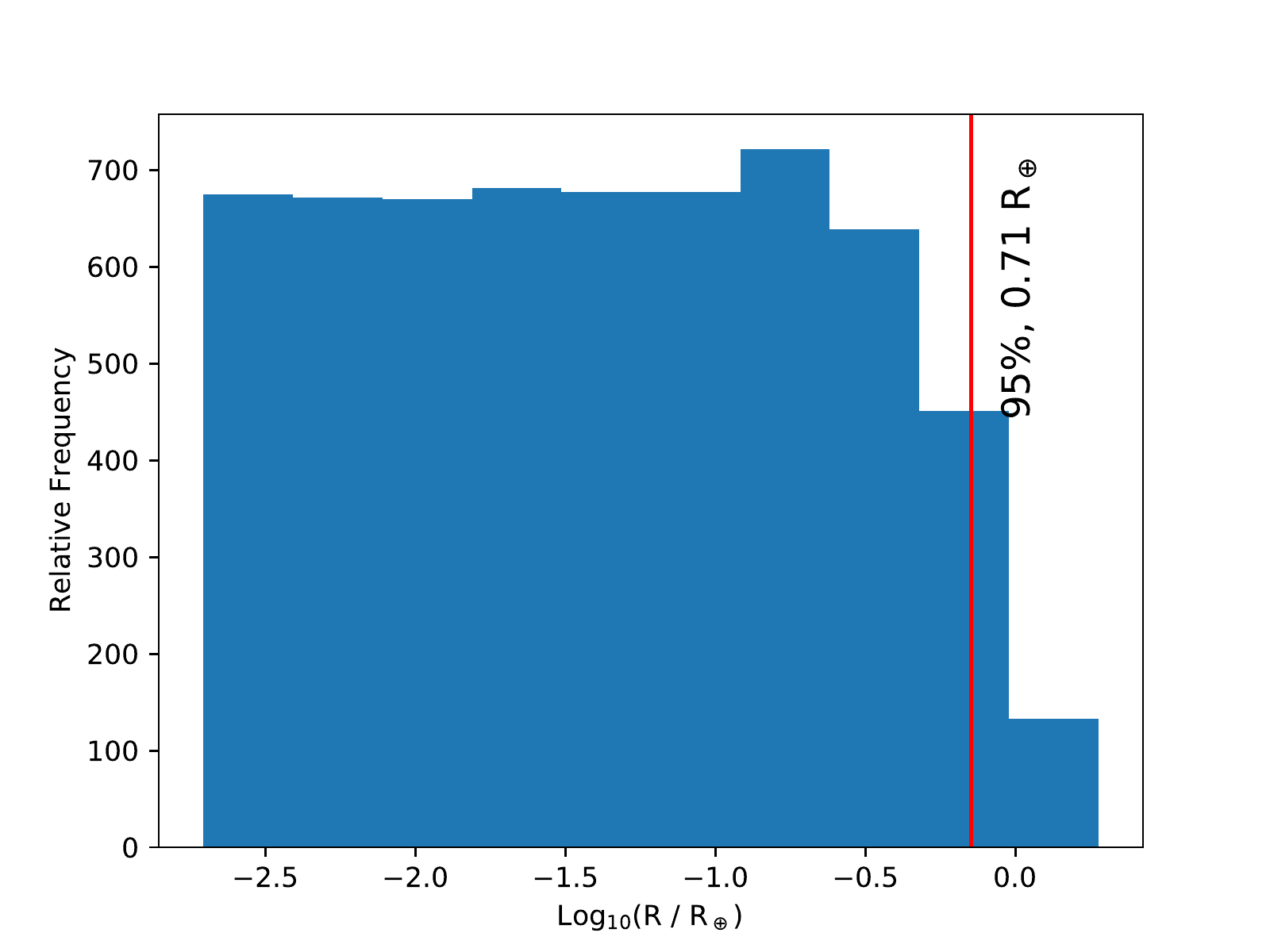}{0.49\textwidth}{Posterior Distribution for Planet Radius}
          }
\caption{The g-band photometry from UT 2020-02-20 puts upper limits on the size of the underlying disintegrating planet.
The minimum radius from \citet{sanchis-ojedak2-22} of 2.5 R$_\oplus$ is shown as a green lightcurve, as well as the 95\% upper limit of \maxRadiusEarths.
The lightcurve is fit with \texttt{exoplanet} \citep{foreman-mackey2021exoplanet} with a Gaussian process (orange 95\% contours) as well as binned data for 10 minute bins (orange error bars).
{\it Right:} Posterior distribution of the planet radius, with the 95\% upper limit highlighted.
The lower limit on the planet size prior comes from adding all of the observed mass loss together into a minimum size rocky planet (13 km), as discussed in Section \ref{sec:planetSize}.}\label{fig:photFeb20gpFits}
\end{figure*}

The shallow transit of \shab\ on UT 2020-02-20 allows us to put an upper limit on the size of the disintegrating planet.
We note that the faintness of the system made it difficult to put accurate constraints on the size of the planet from individual transits with $K2$, with a transit depth constrained to be less than 0.14\% for selected shallow transits \citep{sanchis-ojedak2-22}.
We use the $g$ band lightcurve because it has much higher precision and smaller systematics than the $K$ band.
We fit the lightcurve using \texttt{exoplanet} \citep{foreman-mackey2021exoplanet} and the dependencies therein (see Software) to find the posterior distribution of the planet size.

We start by defining a prior on the planet size by estimating its minimum possible radius.
\shab\ has been disintegrating for at least the 6 years it has been observed.
If we multiply a 6 year lifetime of \sha\ by $\dot{M}$, the minimum mass is 5$\times 10^{16}$ kg.
Converting this into a rocky body with a density of 5,500 kg m$^{-3}$, this value translates to a minimum radius of \minRadius\ at the start the \deleted{the }K2 mission.
It is very unlikely that the K2 mission began observing \shab\ just as it started to disintegrate nor that it was completely destroyed by UT 2020-02-20, so the true lifetime and radius of the planet are very likely be much larger.
However, this allows us to set a prior on the planet size that is not arbitrarily small.
We set a uniform prior in logarithmic space between \minRadius\ and 10\% the stellar radius (ie. \minRadiusEarths\ to 6 R$_\oplus$).
We set a \texttt{pmc3} Beta function prior with $\alpha$=1.0 and $\beta$=3.0 on the impact parameter \citep{salvatier2016pymc3} to favor a low impact parameter because a grazing transit is both relatively unlikely by chance and would require a large vertical ejection of dust grains to produce the observed transit depths of up to $\sim$1.3\%.
We also do not detect any extremely short timescale events visible in the K2 or ground-based lightcurves as would be expected for a grazing transit.
We fix the planet orbit as circular given the likely fast circularization timescales of planets at short orbital periods.

As with many ground-based lightcurves, there are time-correlated noise sources visible on all nights of observation.
We model these correlations as a Gaussian process where the covariance can be modeled with a single simple harmonic oscillator term using \texttt{celerite2} \citep{foreman-mackey2018celerite,foreman-mackey2017fastScalableGPs}.
We adopt a log-normal prior of the timescale of the correlations with a timescale of 0.02 days.
We include as free parameters in the fit the mean of the lightcurve, the planet radius, the impact parameter, the Gaussian process errors and timescales.
We note that the Gaussian process allows for a wider range of planet radii than a simpler lightcurve fit because it allows for the possibility that a lightcurve fluctuation could artificially lower the transit depth.

The posterior distribution of the Gaussian-process lightcurves can be seen in Figure \ref{fig:photFeb20gpFits}.
For clarity, we show the lightcurve with 10 minute long bins but the full lightcurve is used in the Gaussian process likelihood function.
It is clear from the lightcurve that the upper limit of 2.5 $R_\oplus$ from \citet{sanchis-ojedak2-22} is \replaced{imporoved}{improved} by the LBT g-band lightcurve.
The 95\% upper limit in the posterior is \maxRadiusEarths\ for the planet radius.
This improves on the previous upper limit \citep{sanchis-ojedak2-22} by a factor of \maxRadiusImprovementFactor.
\shab\ therefore still has a range of possible radii from \minRadius\ to \maxRadius.

\subsection{Planet Lifetime Limits}\label{sec:lifetime}

\begin{figure*}
\gridline{\fig{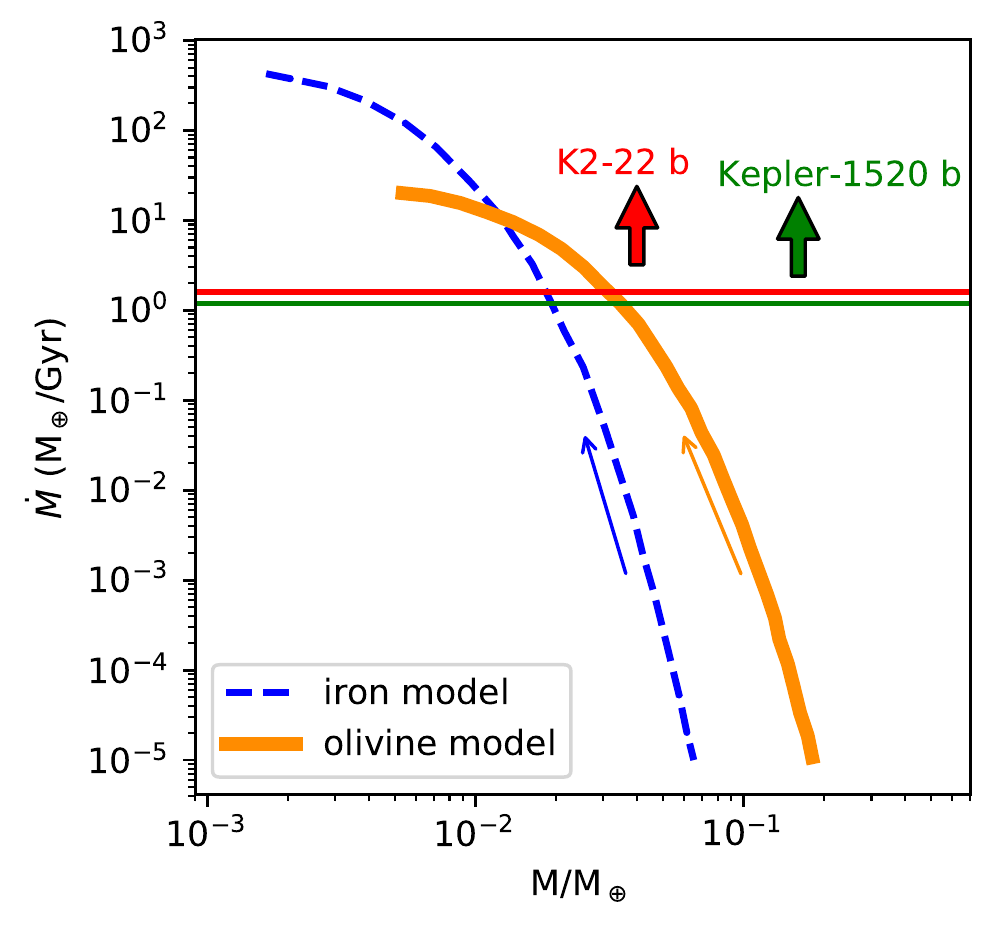}{0.48\textwidth}{}
\fig{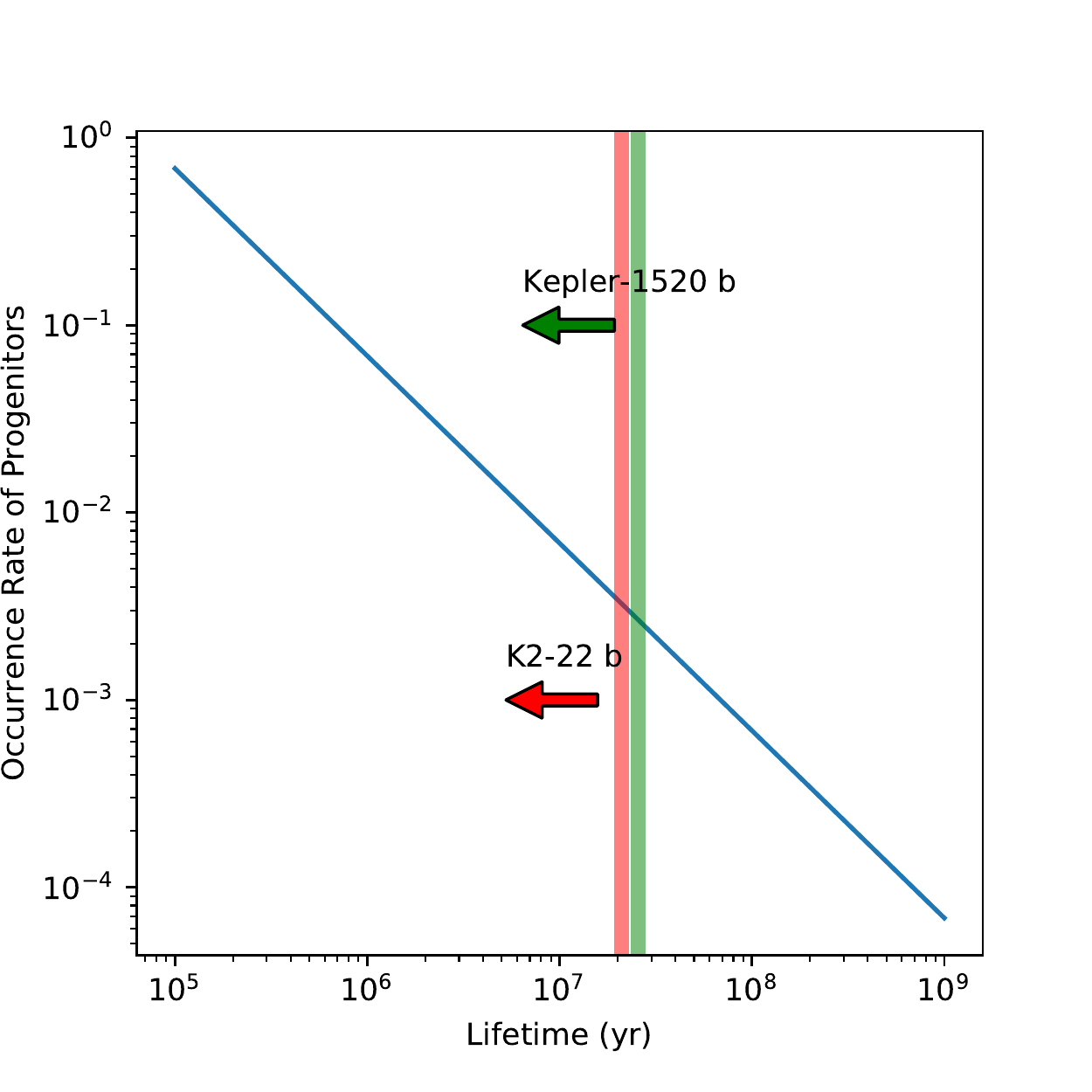}{0.48\textwidth}{}}
\caption{{\it Left:} The mass loss rate trajectories as a function of mass from hydrodynamical models \citep{perez-becker} are indicated for an olivine composition planet and an iron composition planet in orange and blue, respectively, with arrows indicating the direction of evolution.
The high mass loss rates for \shab\ and \shbb\ (with lower limits shown as horizontal red and green lines, respectively) imply that the underlying planets are in their final catastrophic state of evolution.
{\it Right:} Given the short lifetime of the systems (with upper estimates shown as vertical lines for \shab\ and \shbb, respectively), we expect that short-period small progenitors may be common in more than \occurrenceProgenitors\ of systems, using Equation \ref{eq:fobsfprog}.
}\label{fig:mdotEvolution}
\end{figure*}

Combining our mass loss rate and upper size limit on the planet indicates a short lifetime for the system.
If we assume it has a density of 5,500 kg m$^{-3}$, the \maxRadiusEarths\ upper size limit implies a mass of \maxMassEcomp\ or smaller.
Combining this with the $\dot{M}$ from above, the lifetime of the planet is less than \lifetimeE.
If \shab\ is pure iron, the density could be similar to the $\epsilon$ phase of iron, or 8,300 kg m$^{-3}$ \citep{seager2007massRadius} and the lifetime estimate increases to \lifetimeIron.

The high mass loss rate observed for \shab\ and \shbb\ implies they are near the end of their lifetime.
In Figure \ref{fig:mdotEvolution}, we show the mass loss rate from the hydrodynamical model of \shbb\ from \citet{perez-becker}.
The mass loss rate increases as the mass of the underlying planet decreases until it reaches the free streaming limit.
With this model, the mass of the underlying planet is predicted to be $\lesssim 0.03$~\mearth\ for an olivine composition.
Putting this together with the $\dot{M} \approx$~\mdotMePerGyr, the expected lifetime of the planet is $\lesssim$ \lifetimeModel.
We also use the transit depth, particle size distribution of $\gtrsim$0.5~\micron~ from \citet{schlawin2016kic1255} to estimate the $\dot{M}$ for \shbb\ as 1~\mearth/Gyr for comparison.

We use the planet lifetime calculations to estimate the occurrence rate of progenitors following \citet{perez-becker}.
We begin by estimating $f_{observed}$, the fraction of stars with observed disintegrating planets, using the ones from the main Kepler mission, which as a better characterized sample.
There were two periodic disintegrating systems observed from the Kepler main mission: Kepler-1520~b \citep{rappaport} and KOI-2700~b \citep{rappaport2014KOI2700}.
From the catalog of \citet{berger2018keplerRadiiGaiiDR2}, there are 116,637 FGK dwarf stars that not marked as binaries or evolved \citep{kunimoto2020occurrenceRateFGK}.
Therefore, we estimate $f_{observed}$ = 2/116,537.
Following \citet{perez-becker}, we solve
\begin{equation}\label{eq:fobsfprog}
f_{observed} \sim f_{prog} f_{transit} f_{evap},
\end{equation}
where $f_{prog}$ is the occurrence rate of progenitors, $f_{transit}$ is the geometric probability of transit $\sim$ 0.25 and $f_{evap}$ is the fraction of a system's lifetime where the planet is undergoing observable evaporation.
The age of \sha\ is $\gtrsim$1 Gyr \citep{sanchis-ojedak2-22} so we estimate $f_{evap}$ as the \lifetimeModel/1 Gyr.
Solving Equation \ref{eq:fobsfprog} for $f_{prog}$, we find that more than \occurrenceProgenitors\ of systems harbor a progenitor to a disintegrating planet.
High precision future observatories may uncover new short period planets with small Mercury size or smaller bodies if they are this common.

\section{Conclusions}\label{sec:conclusions}
We observed the K2-22 b system with high precision simultaneous multi-wavelength photometry with the LBT and the 16-inch Hereford observatory.
We observed transits on three nights with medium to shallow transit depths as compared to the average from the K2 mission.
We measured the transit depth for a medium-depth event and find that the transmission spectrum is either flat or rising slightly toward long wavelengths.
This transmission spectrum allows us to constrain the dust particle size distribution.
We adopt a log-normal distribution of particle sizes for a variety of single compositions.
For a Mg$_2$SiO$_4$ (forsterite) composition the median particle size of the distribution is constrained to be larger than 0.8~\micron.
This value can vary with composition from 0.5~\micron\ for Fe (metallic iron) to 1.0~\micron\ for SiO$_2$ (quartz).
This is just a lower limit set by the longest wavelength (2.16~$\mu$m), so the median particle size could be larger for all compositions.
Observations with JWST NIRSpec and MIRI can place valuable new constraints on the particle size distribution as well as the composition through mid-infrared solid state features \citep{bodman2018disintegratingInteriors,okuya2020disintegratingCompositionSPICA}.

The large particle sizes, larger surface area and the short time for the dust tail to be cleared and recreated (1 orbital period or 0.4 days) implies a large mass loss rate for the planet of $\dot{M} \gtrsim$ \mdot\ or equivalently \mdotMePerGyr.
This is just a lower limit for the mass loss rate because the particles could be significantly larger than 0.5~\micron.
Hydrodynamic escape of a metal-rich atmosphere surrounding the planet \citep{perez-becker} predicts that $10^8$~kg/s mass loss rates are possible in a planet's final catastrophic phase of disintegration.
\deleted{We note that this is similar to the mass of aerosols from the Krakatoa 1883 volcanic eruption \citep{rampino1982historicVolcanicEruptions} every 100 seconds.}
\added{One other mechanism that has been discussed is explosive volcanism \citep{rappaport}.
We note that the observed minimum $\dot{M}$ is orders of magnitude higher than has been observed for dusty mass loss in the present-day Solar System \citep[$\dot{M}_{Io} \approx$ 1000 kg/s][]{geissler2003volanicActivityIo}, so it would require much more vigorous activity for \shab's mass loss to be produced by volcanism.}
\deleted{Hydrodynamic escape of a metal-rich atmosphere surrounding the planet \citep{perez-becker}, however, predicts that $10^8$~kg/s mass loss rates are possible in a planet's final catastrophic phase of disintegration.}

We also used the very shallow transit depth of UT 2020-02-20 to put new upper limits on the size of the underlying disintegrating planet.
We used the mass loss rate and the fact that disintegration has been observed for at least 6 years to set the minimum radius prior.
Our upper limit is \maxRadiusEarths\ compared to the previous value of 2.5~R$_\oplus$.
The range of allowed radii is \minRadius\ to \maxRadius.

Combining \replaced{this}{the $\dot{M}$} with our radius constraints, the characteristic lifetime of the planet (M/$\dot{M}$) is $\lesssim$~\lifetimeIron.
\added{It is} $\lesssim$~\lifetimeModel\ when assuming the mass from a hydrodynamic model \citep{perez-becker}.
This relatively short lifetime indicates we are observing a short catastrophic phase of the planet disintegration.
If \shab\ and another disintegrating planet \shbb\ have lifetimes this short, it implies that their progenitors are common.
We expect that high precision results from future transit surveys may uncover that short period Mercury and smaller-size progenitors occur in more than \occurrenceProgenitors\ of systems.

\acknowledgments

\section*{acknowledgements}
Funding for \deleted{the }E Schlawin is provided by NASA Goddard Spaceflight Center.
This research has made use of NASA's Astrophysics Data System Bibliographic Services.
We are very grateful to Bruce Gary for providing lead-up photometry as well as the simultaneous unfiltered CCD photometry and his feedback on this work.
We wish to thank Olga Kuhn, Jennifer Power, Andrew Cardwell and Michelle Edwards for invaluable work preparing and observing in the LBT mixed-mode time series.
We acknowledge support from the Earths in Other Solar Systems Project (EOS, PI: Apai), grant No. 3013511 sponsored by NASA. The results reported herein benefited from collaborations and/or information exchange within NASA’s Nexus for Exoplanet System Science (NExSS) research coordination
network sponsored by NASA’s Science Mission Directorate.
Thank you to Graham Lee and Jake Taylor for compiling and sharing the complex indexes of refraction for the dust particles used in this work.
\added{We thank the anonymous reviewer for taking the time to review this work and provide helpful clarifications and corrections.}

%

\vspace{5mm}
\facilities{Kepler, LBT(LBCB and LUCI2)}


\software{astropy \citep{astropy2013}, 
          \texttt{emcee} \citep{foreman-mackey2013emcee},
          \texttt{photutils v0.3} \citep{bradley2016photutilsv0p3},
          \texttt{ccdproc} \citep{craig2015ccdproc}
          \texttt{matplotlib} \citep{Hunter2007matplotlib},
          \texttt{numpy} \citep{vanderWalt2011numpy},
          \texttt{scipy} \citep{virtanen2020scipy},
          \texttt{pymc3}, \citep{salvatier2016pymc3},
          \texttt{starry} \citep{luger2019starry},
          \texttt{arviz} , \citep{kumar2019arviz},
          \texttt{theano}, \citep{theano2016theano},
          \texttt{exoplanet} \citep{foreman-mackey2021exoplanet,agol2020exoplanetAnalytic},
          \texttt{celerite} \citep{foreman-mackey2018celerite,foreman-mackey2017fastScalableGPs}
           }



\appendix

\section{Flat Field Calibration Image Creation}\label{sec:flatFieldCreation}

\begin{figure*}
\gridline{
          \fig{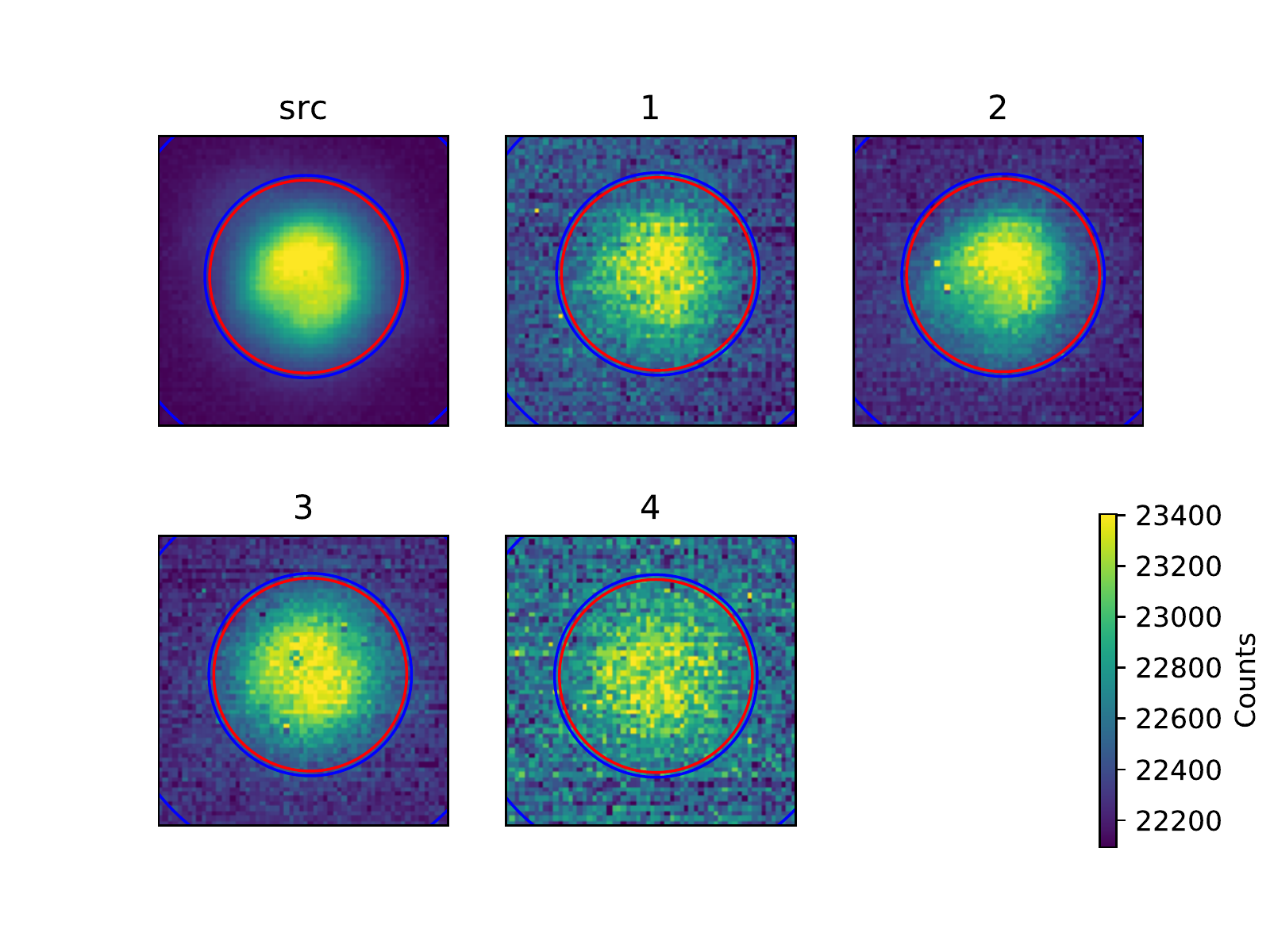}{0.49\textwidth}{Image with Original Flat Field}
          \fig{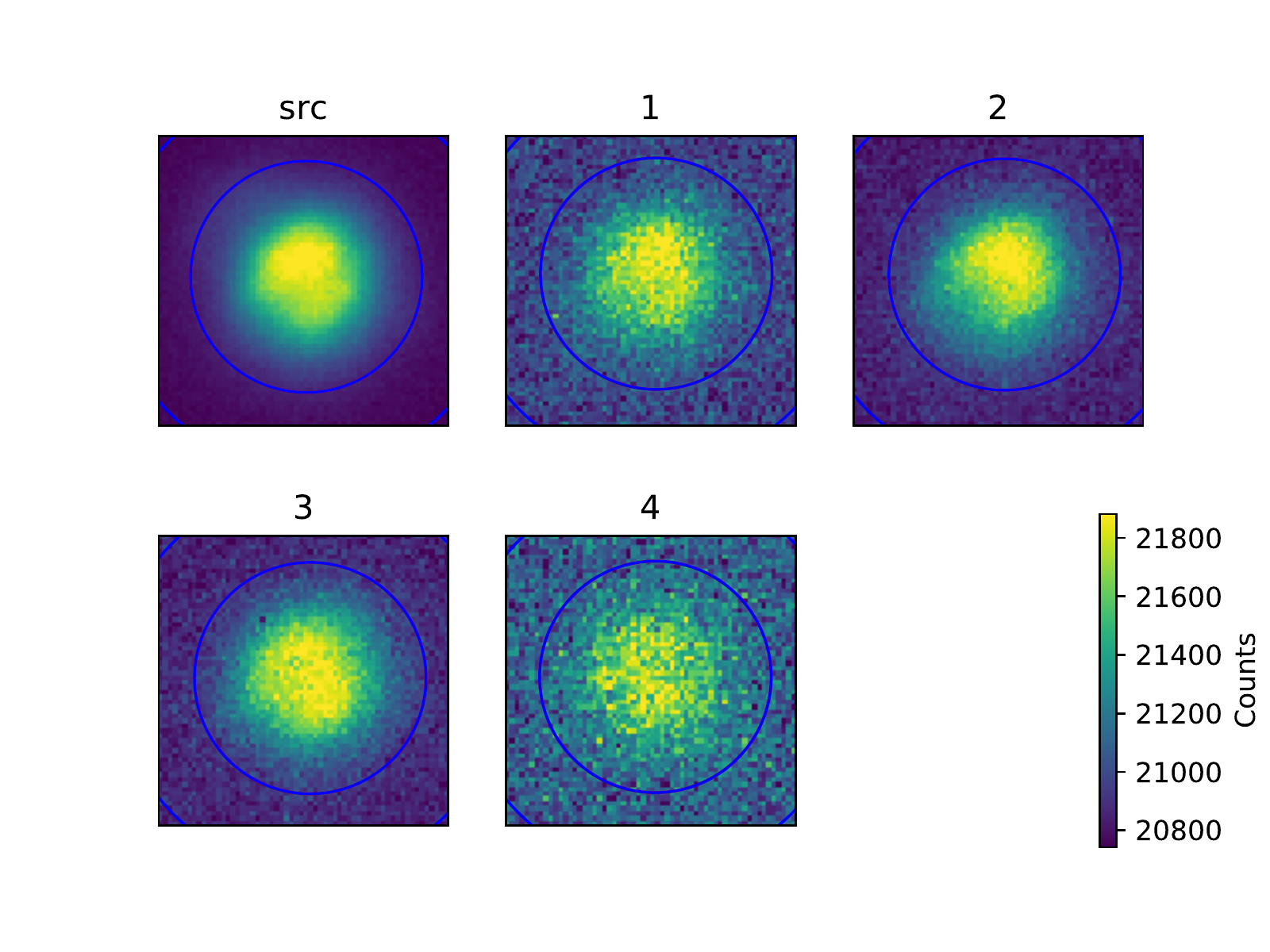}{0.49\textwidth}{Image with Final Flat Field}
          }
\caption{Postage stamp cutouts near the \sha\ and reference stars used to remove common-mode systematics for the infrared LUCI2 HgCdTe detector.
Applying the modified flat field from Equation \ref{eq:modFlat} reduces the horizontal striping (most visible in star 4) in the science images.
Additionally, replacing the bad pixels in the flat field calibration image reduces the number of artifacts in the science data.
}\label{fig:starPostageStamps}
\end{figure*}

The dome flat field was observed at high and low lamp illumination, so we made average flat fields of all exposures ($f_{high}$ and $f_{low}$) for the high and low illumination levels.
We expected to subtracted the two values, ie.
\begin{equation}
f_{original} =  \frac{f_{high} - f_{low}}{Med(f_{high} - f_{low})},
\end{equation}
where Med() is the median of an image to normalize it to 1.0.
However, it was necessary to add a further correction proportional to the ratio of the high and low illumination flat field.
We used the following prescription for the flat field:
\begin{equation}\label{eq:modFlat}
f_{final} = \left( f_{high} - f_{low} \right) \times \left( -2.0 (f_{ratio} - 1.0) + 1.0 \right),
\end{equation}
where
\begin{equation}
f_{ratio} = \frac{f_{high} / f_{low}}{\mathrm{Med}\left( f_{high} / f_{low} \right)}.
\end{equation}
As shown in Figure \ref{fig:starPostageStamps}, applying the modified flat field has less pronounced horizontal stripes than the unmodified original flat field.
This reduces the position-dependence of the time series.

After the flat field correction, there are many bad pixels in $f_{ratio}$ near the target and reference stars that can affect photometry.
We identified the bad pixels from $f_{ratio}$: any points below 0.93 or 1.25 from the median were marked as bad pixels.
Values this far from the median ratio indicated that the pixels responded highly non-linearly or not at all with respect to the median pixel response.
Also, any points with $f_{high} - f_{low}$ $<$ 1600 counts were marked as bad pixels because they did not respond to the differential flux.
The bad pixels in $f_{ratio}$ are replaced with 1.0.
The resulting flat field produces fewer bad pixels in the final reduced images, as seen in Figure \ref{fig:starPostageStamps}.

\bibliographystyle{apj}
\bibliography{this_biblio}



\end{document}